# Structural and Magnetic Properties of Small Symmetrical and Asymmetrical sized Fullerene Dimers

Sandeep Kaur[1,a], Amrish Sharma[1,b], Hitesh Sharma[2,c] and Isha Mudahar[3,*]

[1]Department of Physics, Punjabi University, Patiala, India

[2]Department of Applied Sciences, IKG Punjab Technical University, Kapurthala, India

[3]Department of Basic and Applied Sciences, Punjabi University, Patiala, India

[a]sandeep_rs16@pbi.ac.in,

[b]amrish99@gmail.com,

[c]hitesh@ptu.ac.in,

[*]dr.ishamudahar@gmail.com, M: +918146992328

**ABSTRACT;** Magnetism in carbon nanostructures is of high scientific interest, which could lead to novel magnetic materials. The magnetic properties of symmetrical and asymmetrical sized small fullerene dimers ($C_n$ for n≤50) have been investigated using spin polarized density functional theory. The interaction energies depict that small fullerene cages form stable dimer structures and symmetrical sized fullerene dimers are found more stable than asymmetrical sized dimers. The dimerization of fullerene cages in different modes leads to change in their magnetic properties. The non-magnetic fullerene cages become magnetic after formation of dimer ($C_{20}$-$C_{20}$, $C_{24}$-$C_{24}$, $C_{32}$-$C_{32}$, $C_{40}$-$C_{40}$, $C_{20}$-$C_{24}$, $C_{40}$-$C_{44}$ and $C_{44}$-$C_{50}$), whereas the magnetism of magnetic fullerenes is enhanced or lowered after dimerization ($C_{28}$-$C_{28}$ $C_{36}$-$C_{36}$, $C_{24}$-$C_{28}$, $C_{28}$-$C_{32}$, $C_{32}$-$C_{36}$ and $C_{36}$-$C_{40}$). The individual cages of dimer structures show ferromagnetic interactions amongst them and resultant magnetic moment strongly depends on the type of inter-connecting bonds. The magnetism may also be explained based on distortion of carbon cages and change in the density of states (DOS) in dimer configuration. The calculations presented show strong possibility of experimental synthesis of small fullerene based magnetic dimers.

**Keywords:** Carbon nanostructures, Small fullerenes, Density functional theory, Magnetism.

1.  INTRODUCTION;

The non-IPR (Isolated Pentagon Rule) fullerenes or small fullerenes ($C_n$, n<60) are interesting to study as they exhibit significant structural, electronic and magnetic properties, owing to their high curvature and fused adjacent pentagons [1-5]. The fullerenes have been widely studied in recent years and they have been explored for emerging potential applications in various areas of research such as nano-electronics, molecular devices, spin-electronics etc. [6-8]. The applications of these fullerenes in the field of chemical catalysis [9] and pharmaceutics [10] have also become important by virtue of their particular properties like high chemical reactivity and small diameter.

In past, carbon based systems have become increasingly interesting due to their significant magnetic properties and they can be considered as a possible magnetic materials [11]. The origin of carbon based ferromagnetism has been reported due to the dislocations, vacancies and impurity atoms [12]. Till date, various attempts have been made to study the magnetic properties of small and larger fullerene cages [13-15]. Synthesis of ferromagnetic polymerized fullerenes has been treated by photo assisted oxidation, which show magnetization of order $10^{-3}$ $\mu_B$ per $C_{60}$ [16]. There is an introduction of strong magnetism in the fullerene cages when they are endohedrally doped [14, 15, 17]. When the transition metals (TM) are encapsulated inside small carbon cages, the magnetic behavior of small cages is altered, which furnish a novel possibility to control the magnetic properties of carbon systems [17]. A theoretical and experimental study on carbon clusters show that their magnetic moment can be significantly enhanced by appropriately choosing their size, geometry and composition [18].

Apart from this, the fullerene cages also have ability to form dimer structures. The formation of $C_{60}$ dimer has been confirmed by mechanochemical synthesis experimentally using high speed vibration miling (HSVM) technique [19]. It is found that the stability of $C_{60}$ dimer is comparable to that of two $C_{60}$ molecules and the results are in agreement with theoretical calculations [20]. The dimer structures of fullerene cages can also be obtained through combination of either functionlized fullerenes or bifunctional cycloadditions to $C_{60}$ fullerene cages [21]. Thermal reactions of $C_{60}/C_{60}O/C_{60}O_2$ system lead to the formation of other dimeric fullerene derivatives such as $C_{120}O_2$ and $C_{119}$ [22]. The sulfur containing fullerene derivative $C_{120}OS$ has been formed by thermolysis of $C_{120}O$ in the presence of sulfur [23]. Scanning tunneling microscopy (STM) and low-energy electron diffraction (LEED) studies on polymerized $C_{60}$ studies reveal that annealing of the electron-beam-modified surfaces restores the fullerene lattice [24]. The heating of dimerized $C_{60}$ structures at high temperature restores the fullerenes to their pristine state [18, 24]. Some more experiments have been performed to produce the carbon bridged dimers like $C_{121}$ and $C_{122}$, which could be used as the basic units of fullerene chain structures [25-29].

The existence of $C_{60}$ dimer was also reported theoretically in metastable phases of $MC_{60}$ (M = K, Rb, Cs) [30] and the dimerization can occur in different phases like dumb-bell [31], peanut and capped nanotubes [32]. The conductance of the dimer can be tuned with doping as a result of which it become more versatile in molecular electronics [33]. The magnetic properties of $C_{60}$ dimer indicate the presence of strong magnetic field at cage centers of the dimer and the addition of C-bridges change the behavior of magnetic field. The fullerene dimers connected through BN hexagons alter the behavior of magnetic field inside the cages [34]. Ab-initio calculations show that unpaired electrons of $C_{59}N$ are delocalized over $C_{60}$ molecule in $C_{59}N$-$C_{60}$ hetrodimer [33]. Nucleus Independent Chemical Shift (NICS) and Nuclear Magnetic Resonance (NMR) studies have reported the magnetic properties of $(C_{58}BN)_2$ dimer and the results reveal that the dimerization of fullerene cages causes the major changes in magnetic properties [35]. $C_{36}$ cage can also form strong inter cage bonds and may be a hexavalent building block for fullerene compounds like dimers and polymers [36]. A computational study [37] on $C_{36}$ shows the dimerization of the cage, but the experimental existence of dimer is not yet confirmed. The endohedral derivatives of $C_{36}$ dimer are also expected to exist with their unique properties, making it useful for molecular devices [38]. However, no systematic study on small fullerene dimers has been reported yet. Because of the limited study on dimer structures, we are reporting for first time a systematic study on symmetrical and asymmetrical sized small fullerene dimers. The motivation for considering asymmetrical sized dimers comes from the experimental existence of carbon nanobuds, which are also a combination of asymmetrical sized nanostructures [39].

In the present work, we employed the first principle calculations on the dimers of small fullerene cages ($C_n$, n≤50) based on density functional theory. Since $C_{60}$ dimer has significant properties, so it is interesting to study the interaction between small fullerene cages, which adds a new dimension to dimer properties.

## 2. COMPUTATIONAL DETAILS;

All the calculations were performed using Spanish Initiative for electronic simulation with thousands of atoms (SIESTA) computational code, which is based on density functional theory [40]. The Perdew, Burkey and Ernzerhof (PBE) functional combined with double-ζ polarized basis set were used for the geometry optimizations [41]. Kleinman – Bylander form of non-local norm conserving pseudo potentials are used to describe core electrons [42] while numerical pseudoatomic orbitals of the Sankey – Niklewski type [43] are used to represent the valence electrons. The energy shift parameter is defined by range ≈ 150 – 350 meV to describe the size of pseudoatomic orbitals. The fineness of a finite grid is defined by Mesh cut-off, whose value lie within the range 150Ry – 250Ry. In order to obtain ground state properties, minimization of total energy of the system has been executed. The residual forces of the system are relaxed up to 0.004 eV/Ang. The energy eigen values have been plotted in order to give density of states (DOS) spectra.

Test calculations were performed on small fullerenes ($C_n$, n < 60) to check the accuracy of our results. The geometrical parameters for small fullerenes are calculated and shown in Table 1 and the results are in agreement with known experimental and theoretical results [3, 44]. We found that our calculated values follow the same pattern and are in agreement with a recent study which employs tight binding method [45]. Since our group has already study the carbon based systems so the parameters have been checked [17, 46-47].

**Table 1.** Average Diameter, Average Bond Distance and HOMO-LUMO gaps of small fullerene cages.

| Fullerene Cage | $D_{av}$ (Å) | Average Bond Distance (Å) | HOMO-LUMO Gap (eV) |
|---|---|---|---|
| $C_{20}$ | 4.14 | 1.48 | 0.75 |
| $C_{28}$ | 4.83 | 1.49 | 0.33 |
| $C_{32}$ | 5.25 | 1.46 | 1.41 |
| $C_{36}$ | 5.60 | 1.46 | 0.43 |

## 3. RESULTS AND DISCUSSION;

### 3.1 Structural Properties:

We have optimized the ground state structures of symmetrical and asymmetrical sized small fullerene dimers using method described in computational details. The ground state geometries of small fullerene cages are considered to form the dimer structures. In dimeric pattern, the small carbon cages can connect through four possible modes, i.e. (a) point-point mode forming a [1+1] dimer with C-C bond between two cages, (b) side-side mode forming a [2+2] dimer with a 2-fold bond, (c) face-face mode forming a [5+5] dimer with a 5-fold bond between two pentagonal rings, and (d) face-face mode forming a [6+6] dimer with a 6-fold bond between two hexagonal rings as shown in Fig.1. Each dimer structure with these four configurations is optimized and the structure which has minimum total energy is the most stable isomer.

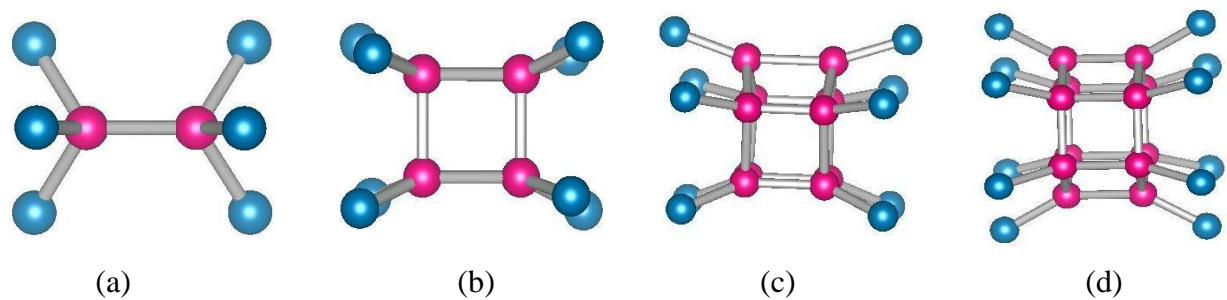

(a)          (b)          (c)          (d)

**Fig. 1**. Four possible configurations through which fullerene cages connect (a) [1+1] point-point mode, (b) [2+2] side-side mode, (c) [5+5] face-face mode between two pentagons and (d) [6+6] face-face mode between two hexagons.

### 3.1.1 Symmetrical Sized Dimers;

Symmetrical sized dimers $C_{20}$-$C_{20}$, $C_{24}$-$C_{24}$, $C_{28}$-$C_{28}$, $C_{32}$-$C_{32}$, $C_{36}$-$C_{36}$, $C_{40}$-$C_{40}$, $C_{44}$-$C_{44}$ and $C_{50}$-$C_{50}$ have been investigated in detail.

$C_{20}$ with $I_h$ symmetry forms $C_{20}$-$C_{20}$ dimer structure in [2+2] side-side mode as the most stable geometry. We have also calculated the relative energy differences ($E_r$) of all the configurations to check for any isomeric structures. In $C_{20}$-$C_{20}$ dimer, [5+5] and [1+1] modes have $E_r$ w.r.t. [2+2] configuration of the order of 0.382 eV and 1.27 eV respectively, which clearly shows that [2+2] mode has the highest stability. The average diameter ($D_{av}$ = 4.14 Å) of cages remains almost same as that of single $C_{20}$ cage. The bond lengths lie in the range 1.38 – 1.40 Å and the connecting bond length between the cages is 1.55 Å as shown in Table 2.

We have calculated the interaction energies (see Table 2) of all the possible structures using the following expression

$$\Delta E = E_{total}(C_n － C_n) － 2E(C_n)$$

where $E_{total}(C_n - C_n)$ and $E(C_n)$ are the total energies of the dimer and individual cages respectively. The negative values of ΔE indicate the greater stabilities of the dimer while the positive values show that the dimer is less stable, so the dimers with negative ΔE have a possibility to be formed experimentally. $C_{20}$-$C_{20}$ dimer with [2+2] side-side mode require more energy (-5.43eV) to dissociate into two cages of dimer structure as compared to other configurations (Fig. 1(a) and 1 (c)), which indicates that this configuration is the most favorable one.

**Table-2.** Interaction energy (ΔE), average diameter ($D_{av}$) and connecting bond length for symmetrical sized dimers.

| Dimer | Interaction Energy (eV) | | | | $D_{av}$.(Å) | Connecting Bond Length (Å) | | | |
|---|---|---|---|---|---|---|---|---|---|
| ($C_n - C_n$) | [1+1] | [2+2] | [5+5] | [6+6] | | [1+1] | [2+2] | [5+5] | [6+6] |
| $C_{20} - C_{20}$ | -4.16 | -5.43 | -5.04 | - | 4.17 | 1.51 | 1.55 | 1.61 | - |
| $C_{24} - C_{24}$ | -3.27 | -4.95 | -4.87 | -5.13 | 4.48 | 1.53 | 1.56 | 1.54-1.75 | 1.59 |
| $C_{28} - C_{28}$ | -2.47 | -3.42 | -3.67 | -2.66 | 4.86 | 1.56 | 1.58 | 1.58-1.60 | 1 .62 |
| $C_{32} - C_{32}$ | -1.59 | -2.54 | -0.76 | -0.97 | 5.22 | 1.54 | 1.56 | 1.58-1.62 | 1.59 |
| $C_{36} - C_{36}$ | -2.93 | -2.72 | -3.00 | -2.22 | 5.50 | 1.54 | 1.58 | 1.58-1.60 | 1.59 |
| $C_{40} - C_{40}$ | -1.97 | -2.21 | -0.87 | -1.96 | 5.69 | 1.57 | 1.56 | 1.58-1.65 | 1.58-1.62 |
| $C_{44} - C_{44}$ | -1.76 | -2.59 | -2.42 | -1.33 | 6.00 | 1.51 | 1.58 | 1.55-1.59 | 1.55-1.58 |
| $C_{50} - C_{50}$ | 0.02 | -1.46 | 0.64 | 0.86 | 6.65 | 1.59 | 1.57 | 1.56-1.61 | 1.57-1.60 |

$C_{24}$ with $D_{6d}$ symmetry forms a most stable dimer with [6+6] face-face mode between two hexagons. In $C_{24}$-$C_{24}$ dimer, $E_r$ for [2+2] mode w.r.t. [6+6] mode is 0.176 eV which point towards the stability of [2+2] mode as well. The energy difference for other modes w.r.t. [6+6] mode is comparatively larger. The average diameter of $C_{24}$ cage in isolated form is 4.69 Å, which decreases to 4.48 Å when connected in [6+6] mode of dimer leading to distortion in the structure. The bond lengths vary between 1.40 – 1.55 Å and the connecting bond length between the cages is 1.59 Å. From the values of interaction energy, [6+6] configuration has maximum chances to be formed as compared to other counterparts.

$C_{28}$ cage having $T_d$ symmetry forms a most stable structure with [5+5] face-face mode between two pentagons after their dimerization. [1+1] and [6+6] configurations have $E_r$ w.r.t. [5+5] of the order of 1.20 eV and 1.01 eV respectively. [2+2] configuration has $E_r$ 0.259 eV which indicates that after [5+5] mode, it has the highest probability to be formed. The average diameter (4.83 Å) of $C_{28}$-$C_{28}$ dimer remains identical as compared to the individual cage. The C-C bond lengths vary from 1.40 Å to 1.58 Å in $C_{28}$-$C_{28}$ dimer and the connecting bond lengths lay from 1.58 – 1.60 Å. ΔE values indicate that the dimer structure with [5+5] mode has largest stability amongst all the other modes.

$C_{32}$ ($D_3$), $C_{40}$ ($D_2$), $C_{44}$ ($D_2$) and $C_{50}$ ($D_3$) cages form most stable dimers when connected through [2+2] side-side mode. For $C_{32}$-$C_{32}$ and $C_{50}$-$C_{50}$, the relative energy differences of [1+1], [5+5] and [6+6] modes are quite high w.r.t. [2+2] mode, which further establishes [2+2] as the most stable configuration. For $C_{40}$-$C_{40}$ dimer, $E_r$ value for [1+1] and [6+6] is almost same of the order of ~0.25eV, which shows that these two modes are isoenergetic. In $C_{44}$-$C_{44}$ dimer, [2+2] mode is isoenergetic with [5+5] mode as the relative energy difference between these two modes is 0.07eV. The $D_{av}$ of individual cages do not show much change after dimerization except $C_{44}$-$C_{44}$ dimer, where $D_{av}$ decreases from 6.15 Å to 6.0 Å. The average bond lengths for $C_{32}$-$C_{32}$ and $C_{44}$-$C_{44}$ dimers lie in the range 1.37 Å – 1.53 Å, whereas for $C_{40}$-$C_{40}$ and $C_{50}$-$C_{50}$ dimers the bond lengths vary from 1.38 Å – 1.60 Å. The cages are held together through the connecting bonds of the order of 1.56 Å – 1.58 Å. For $C_{50}$-$C_{50}$ dimer, only [2+2] configuration is energetically favorable as all the other configurations considered show positive ΔE.

$C_{36}$ with $D_{6h}$ symmetry has a most stable dimer structure in [5+5] face-face mode between two pentagonal rings and this configuration is isoenergetic with [1+1] mode having relative energy difference of 0.06 eV. The connecting bond lengths for both configurations are shown in Table 2 and the $D_{av}$ of dimer structure decreases to 5.5 Å w.r.t. single cage (5.6 Å). Interaction energy shows that the dimer with [5+5] mode is most favorable dimer structure between other isomers of $C_{36}$-$C_{36}$ dimer.

The interaction energies for all symmetrical sized dimers show that these dimer structures are energetically favorable and therefore, are likely to be formed. ΔE value decreases from $C_{20}$-$C_{20}$ to $C_{50}$-$C_{50}$ dimer, which shows that the small sized dimers are more likely to be formed as compared to large sized fullerene dimers. The observed structural behavior of symmetrical sized fullerene dimers can be seen from variation in average diameters and connecting bond lengths. In all the dimers except $C_{24}$-$C_{24}$, $C_{28}$-$C_{28}$ and $C_{36}$-$C_{36}$, the most stable configuration with which the fullerene prefers to attach is [2+2]. $C_{24}$-$C_{24}$ and $C_{28}$-$C_{28}$/$C_{36}$-$C_{36}$ dimer shows maximum preference to be formed in [6+6] and [5+5] mode. The average connecting bond length at which the dimers get stable is 1.58 Å. The fullerene have a tendency to settle away from each other as the single C – C bond length is 1.54 Å. In all the dimers formed, the $D_{av}$ decreases w.r.t. the individual cages. However, when the type of inter-cage bonding changes, there is variation in C – C bonds leading to change in the $D_{av}$, which leads to redistribution of charges at localized sites.

### 3.1.2 Asymmetrical Sized Dimers;

We have extended our investigation to study of asymmetrical sized dimers $C_{20}$-$C_{24}$, $C_{24}$-$C_{28}$, $C_{28}$-$C_{32}$, $C_{32}$-$C_{36}$, $C_{36}$-$C_{40}$, $C_{40}$-$C_{44}$ and $C_{44}$-$C_{50}$ using the method described in computational details. $C_{20}$-$C_{24}$, $C_{28}$-$C_{32}$ and $C_{44}$-$C_{50}$ form the most stable dimer with [2+2] side-side mode. The relative energy differences are high for other configurations as compared to [2+2] side-side mode. The average C-C bond lengths vary from 1.37 Å – 1.57 Å for these three dimer structures. There is variation of the order of 0.10 – 0.15 Å in the $D_{av}$ of $C_{20}$-$C_{24}$ and $C_{28}$-$C_{32}$ in comparison to their individual counterparts, whereas for $C_{44}$-$C_{50}$ $D_{av}$ remains almost same as compared to single cages. The connecting bond lengths and interaction energies are tabulated in Table 3 for all the asymmetrical sized dimers. The interaction energy for these dimers is calculated using the following expression

$$\Delta E = E_{total}(C_n - C_m) - E(C_n) - E(C_m)$$

where $E_{total}(C_n - C_m)$, $E(C_n)$ and $E(C_m)$ are the total energies of the dimer and individual cages respectively and are shown in Table 3.

**Table-3.** Interaction energy ($\Delta E$), average diameter ($D_{av}$) and connecting bond length for asymmetrical sized dimers.

| Dimer | Interaction Energy (eV) | | | | $D_{av}$ (Å) | | Connecting Bond Length (Å) | | | |
|---|---|---|---|---|---|---|---|---|---|---|
| $C_n - C_m$ | [1+1] | [2+2] | [5+5] | [6+6] | $C_n$ | $C_m$ | [1+1] | [2+2] | [5+5] | [6+6] |
| $C_{20} - C_{24}$ | -3.45 | -5.18 | -4.93 | - | 4.22 | 4.70 | 1.52 | 1.56 | 1.57-1.65 | - |
| $C_{24} - C_{28}$ | -2.36 | -3.91 | -4.29 | -3.86 | 4.74 | 4.87 | 1.54 | 1.57 | 1.59-1.64 | 1.58-1.62 |
| $C_{28} - C_{32}$ | -2.19 | -2.59 | -2.23 | -1.84 | 4.73 | 5.39 | 1.55 | 1.57 | 1.58-1.61 | 1.58-1.62 |
| $C_{32} - C_{36}$ | -2.33 | -2.20 | -2.00 | -0.27 | 5.34 | 5.50 | 1.54 | 1.57 | 1.57-1.62 | 1.55-1.62 |
| $C_{36} - C_{40}$ | -2.04 | -1.99 | -1.79 | -0.67 | 5.37 | 5.89 | 1.55 | 1.58 | 1.57-1.63 | 1.53-1.69 |
| $C_{40} - C_{44}$ | -1.48 | -1.37 | -1.13 | -1.38 | 5.89 | 6.34 | 1.56 | 1.59 | 1.57-1.62 | 1.58-1.61 |
| $C_{44} - C_{50}$ | -0.67 | -1.46 | -0.78 | -0.06 | 6.13 | 6.61 | 1.57 | 1.57 | 1.58-1.60 | 1.55-1.65 |

Table 3 shows extra stability of [2+2] mode for $C_{20}$-$C_{24}$, $C_{28}$-$C_{32}$ and $C_{44}$-$C_{50}$ due to higher interaction energy. The connecting bond lengths lie in the range 1.56 – 1.59 Å for these three dimers.

In case of $C_{24}$-$C_{28}$ dimer, the most stable dimer forms when pentagonal ring of $C_{24}$ is combined with pentagon face of $C_{28}$ cage i.e. [5+5] face mode. $E_r$ shows that [1+1], [2+2] and [6+6] modes are less stable as compared to [5+5] mode. The average diameter remains almost same for both the individual cages after their dimerization, while the average C – C bond lengths varies between 1.39 – 1.60 eV. The connecting bond lengths vary from 1.59 – 1.64 Å. ΔE value for [5+5] configuration is highest for $C_{24}$-$C_{28}$ among all other possible modes.

For $C_{32}$-$C_{36}$, $C_{36}$-$C_{40}$ and $C_{40}$-$C_{44}$ dimers, the most stable dimer structure has [1+1] point-point mode. In $C_{32}$-$C_{36}$ and $C_{40}$-$C_{44}$ dimers, the relative energy differences are large for [2+2], [5+5] and [6+6] modes w.r.t. [1+1] configuration. For $C_{36}$-$C_{40}$ dimer, [2+2] side-side mode is isoenergetic with [1+1] having energy difference of 0.06 eV. The $D_{av}$ in $C_{32}$-$C_{36}$ dimer shows variation of 0.1 Å w.r.t. single cages. In $C_{36}$-$C_{40}$ dimer, the $D_{av}$ for $C_{36}$ cage decreases form 5.6 Å to 5.37 Å, whereas for $C_{40}$ cage it increases from 5.75 Å to 5.89 Å w.r.t. the individual cages. The $D_{av}$ increases after dimerization for $C_{40}$-$C_{44}$ dimer as compared to their individual counterparts. The connecting bond lengths are of the order of ~1.55Å and the average C – C bond lengths vary from 1.38Å to 1.60Å. Table 2 shows that these dimers have maximum stability to be formed in [1+1] point-point mode.

The interaction energies for asymmetrical sized dimers show similar behavior as that of symmetrical sized dimers. There is an increase in ΔE from $C_{20}$-$C_{24}$ to $C_{44}$-$C_{50}$ dimer, which suggest that the dimers of small size have more chances to form as compared to larger ones. For all asymmetrical sized dimers except $C_{24}$-$C_{28}$, the average connecting bond length is 1.56 Å, which is comparable to single C – C bond length. In comparison to symmetrical sized dimers, the fullerene cages in asymmetrical sized dimers are more closely bound. There is significant variation in the average diameters of asymmetrical sized dimers, which point towards their valuable magnetic properties.

### 3.2 Magnetic Properties;

To study the magnetic properties of small fullerene dimers, spin polarized calculations have been performed on all possible dimer combinations [1+1], [2+2], [5+5] and [6+6]. The HOMO-LUMO gaps for spin up and down electron states, density of states (DOS), total magnetic moments (MM) and localized magnetic moments (MM) (Fig 2) have been calculated for both symmetrical and asymmetrical sized dimers as summarized below.

### 3.2.1 Symmetrical Sized Dimers;

The symmetrical sized dimers with similar size were taken into consideration to analyze their magnetic properties. The HOMO-LUMO energy gaps for electrons with spin up and spin down and total magnetic moments for $C_{20}$, $C_{24}$, $C_{28}$, $C_{32}$, $C_{36}$, $C_{40}$, $C_{44}$, $C_{50}$ with [1+1], [2+2], [5+5] and [6+6] modes are shown in Table 4.

**Table-4**. Total magnetic moments and HOMO-LUMO gaps for symmetrical sized dimers.

| Dimer | Total Magnetic Moment ($\mu_B$) | | | | Dimer | HOMO-LUMO gaps (eV) | | | | | | | |
|---|---|---|---|---|---|---|---|---|---|---|---|---|---|
| $C_n$-$C_n$ | [1+1] | [2+2] | [5+5] | [6+6] | $C_n$-$C_n$ | [1+1] | | [2+2] | | [5+5] | | [6+6] | |
| | | | | | | ↑ | ↓ | ↑ | ↓ | ↑ | ↓ | ↑ | ↓ |
| $C_{20}$-$C_{20}$ | 2.00 | 0.00 | 2.00 | - | $C_{20}$-$C_{20}$ | 0.91 | 1.02 | 1.13 | 1.13 | 1.37 | 0.65 | - | - |
| $C_{24}$-$C_{24}$ | 2.00 | 0.00 | 2.00 | 4.00 | $C_{24}$-$C_{24}$ | 0.72 | 0.34 | 0.59 | 0.59 | 0.79 | 0.47 | 1.60 | 0.62 |
| $C_{28}$-$C_{28}$ | 6.00 | 4.00 | 6.00 | 4.00 | $C_{28}$-$C_{28}$ | 1.52 | 0.81 | 1.79 | 0.43 | 1.94 | 0.81 | 0.34 | 0.65 |
| $C_{32}$-$C_{32}$ | 2.00 | 0.00 | 2.00 | 0.00 | $C_{32}$-$C_{32}$ | 1.85 | 0.37 | 1.31 | 1.31 | 0.60 | 0.51 | 0.75 | 0.75 |
| $C_{36}$-$C_{36}$ | 2.00 | 4.00 | 2.00 | 3.96 | $C_{36}$-$C_{36}$ | 0.92 | 0.74 | 0.71 | 0.73 | 0.93 | 0.60 | 1.03 | 0.96 |
| $C_{40}$-$C_{40}$ | 2.00 | 0.00 | 2.00 | 0.00 | $C_{40}$-$C_{40}$ | 0.58 | 0.49 | 0.67 | 0.67 | 0.48 | 0.18 | 0.94 | 0.94 |
| $C_{44}$-$C_{44}$ | 0.00 | 0.00 | 0.00 | 0.00 | $C_{44}$-$C_{44}$ | 0.57 | 0.57 | 0.66 | 0.66 | 0.55 | 0.55 | 0.72 | 0.72 |
| $C_{50}$-$C_{50}$ | 1.67 | 0.00 | 1.95 | 0.00 | $C_{50}$-$C_{50}$ | 0.45 | 0.84 | 1.17 | 1.17 | 0.54 | 0.46 | 0.97 | 0.97 |

$C_{20}$ fullerene cage with $I_h$ symmetry has zero magnetic moment in isolated form. However, the symmetry and magnetic state of $C_{20}$ has been a point of disagreement in the reported results, which may be due to the accuracy of the electron correlation effects in small fullerenes [48]. $C_{20}$ has shown magnetic to non magnetic transition and vice versa for $I_h$ and $D_{3d}$ symmetries respectively [49]. After dimerization, $C_{20}$-$C_{20}$ dimer in [1+1] and [5+5] modes shows magnetic behavior with total MM of 2.0 $\mu_B$, whereas the most stable configuration with [2+2] mode shows non-magnetic behavior. The local MMs on all individual C atoms for [5+5] mode in $C_{20}$-$C_{20}$ dimer are shown in Fig 2. The major contribution to total MM is contributed by second nearest neighbors (NNs) from the connecting bond atoms, which contributes about 65% of total MM. Atoms at first NN positions and atoms at connecting bridges contribute about 42% and 10% respectively to total MM. The HOMO-LUMO gap of spin up and spin down electrons show distinct pattern in dimer configurations w.r.t. gap in individual building block. The non-magnetic dimer configurations show equal magnitude of HOMO-LUMO gaps for spin up and spin down electron states, whereas magnetic dimers show unequal HOMO-LUMO gaps.

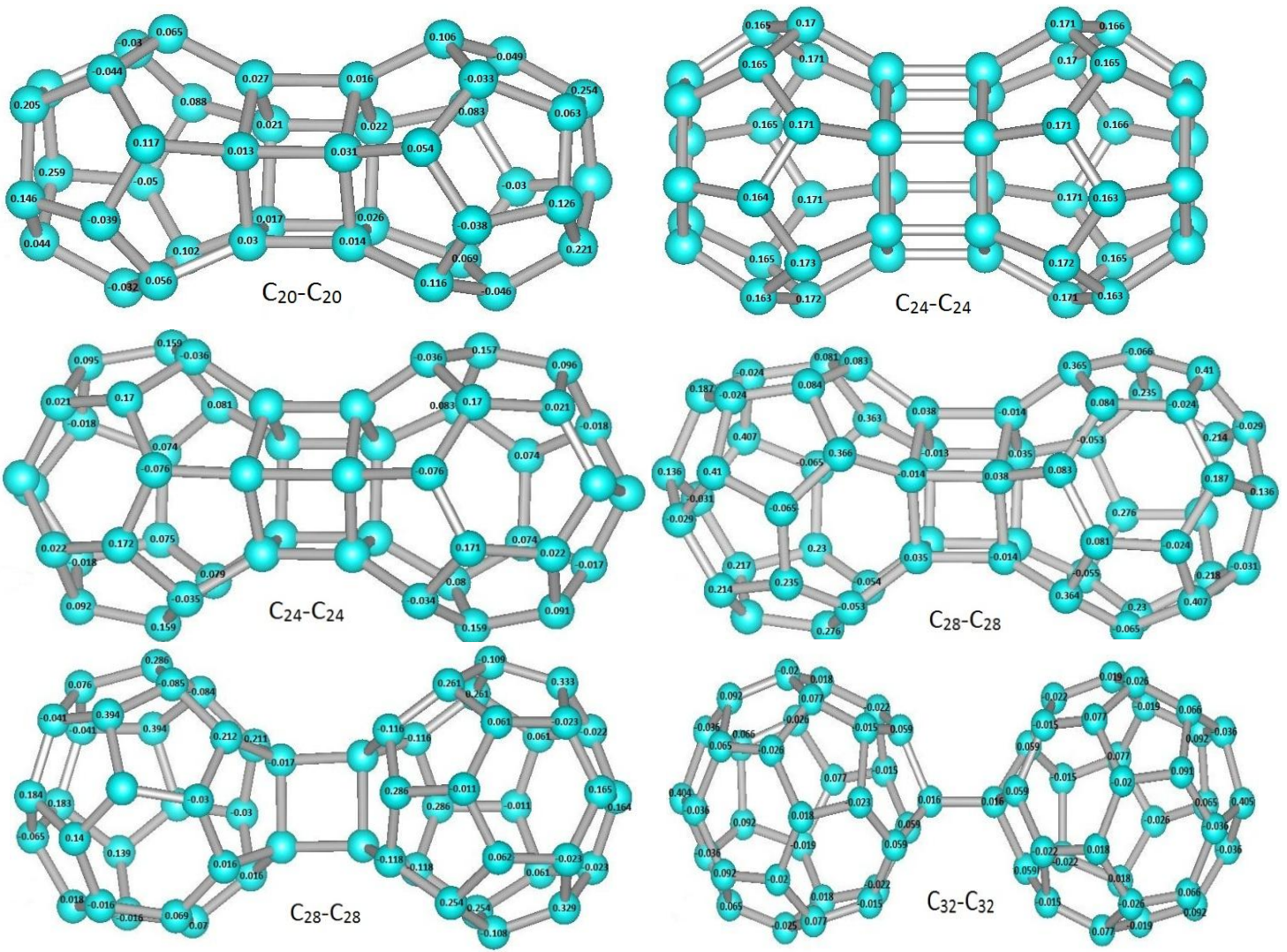

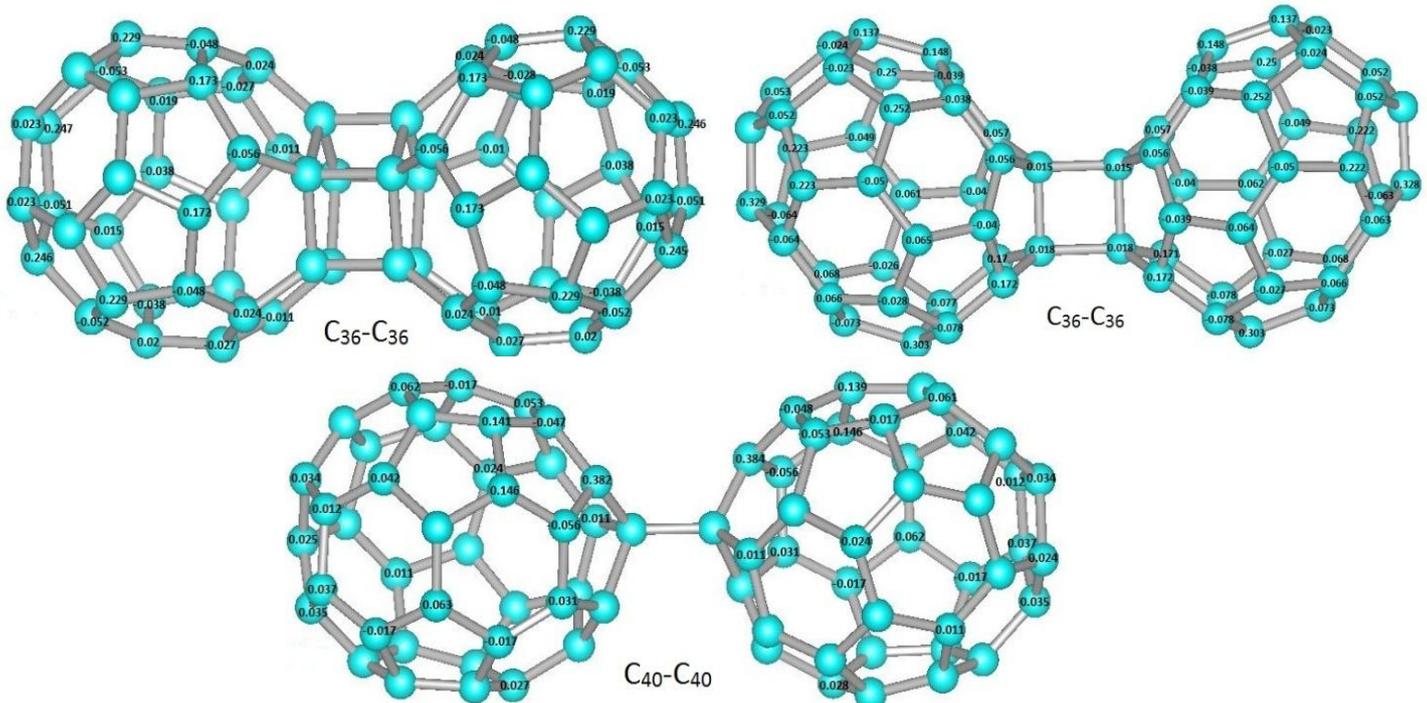

**Fig.2**. Local magnetic moments of symmetrical sized small fullerene dimers in $\mu_B$.

$C_{24}$-$C_{24}$ dimer exhibits high magnetism with total MM of 4.0 $\mu_B$ in [6+6] mode in comparison to isolated $C_{24}$ fullerene which shows 0.0 $\mu_B$ magnetic moment. When connected with [1+1] and [5+5] modes the dimer shows MM of 2.0 $\mu_B$, whereas in [2+2] mode the total MM is 0.0 $\mu_B$. To understand the origin of magnetism, local MMs on all C atoms were calculated. The local MM in [6+6] mode indicate that first and second NNs of both cages contribute ~51% and 49% respectively towards total MM, whereas there is no contribution from connecting bond atoms to total MM. However, in [5+5] mode, 80% of contribution comes from second NNs. The HOMO-LUMO gaps for spin up and spin down states are unequal in [1+1], [5+5] and [6+6] modes showing the magnetic behavior of dimer.

$C_{28}$ fullerene cage has total magnetic moment of 4.0 $\mu_B$ in isolated form. When it forms a most stable dimer in [5+5] mode, total MM increases to 6.0 $\mu_B$. The magnetic moment for [2+2] and [6+6] configurations is 4.0 $\mu_B$, while for [1+1] point-point mode total MM is 6.0 $\mu_B$. The local MM in [5+5] mode show that there is a contribution of 23.5%, 25.8% and 40% for first, second and third NNs respectively towards total MM, whereas in [2+2] mode major contribution comes from the second, third and forth NNs. The connecting bond atoms for both modes show a small contribution of only 1-2% to total MM. All the possible configurations have significant magnetic order in HOMO-LUMO gaps of spin up and spin down states showing their magnetic behavior.

In isolated form $C_{36}$ cage is magnetic with magnetic moment of 2.0 $\mu_B$. The magnetic moment remains same after dimerization of the carbon cage in [5+5] mode. Total MM increases to 4.0 $\mu_B$ for [2+2] and [6+6] mode, while for [1+1] mode it has value 2.0 $\mu_B$. The second, third and fifth NNs form connecting bond atoms make a contribution of 20%, 42% and 48.7% respectively to total MM in [5+5] mode. For [2+2] mode, the major contribution of local MM is 51% for third NN and the remaining local MM are distributed evenly on first, second, fifth and sixth neighbors. The HOMO-LUMO gaps for spin up and spin down states have finite energy difference, which show their magnetic behavior.

As the fullerenes $C_{32}$, $C_{40}$, $C_{44}$ and $C_{50}$ are non-magnetic in their isolated forms and after dimerization in [2+2] mode continuous to remain non-magnetic. However, [1+1] and [5+5] configurations acquire finite magnetic moment after the formation of dimer due to change in inter-cage bonding. In [1+1] and [5+5] modes, $C_{32}$-$C_{32}$ and $C_{40}$-$C_{40}$ dimers acquire finite magnetic moment of 2.0 $\mu_B$, whereas [6+6] mode in both the dimers have zero magnetic moment. The local MMs at C atoms of connecting bonds show no contribution as the magnetic moment is localized away from them. The HOMO-LUMO gaps of spin up and down states are equal for non-magnetic modes [2+2] and [6+6], while the magnetic dimer configurations [1+1] and [5+5] have finite energy difference in their HOMO-LUMO gaps. $C_{44}$-$C_{44}$ dimer is found to be non-magnetic in all the four modes with zero magnetic moment. The HOMO-LUMO gaps for spin up and spin down states are same showing the non-magnetic behavior of the dimer. $C_{50}$-$C_{50}$ dimer is formed only in [2+2] mode and has zero magnetic moment and the dimer has identical HOMO-LUMO gaps for spin up and spin down electronic state.

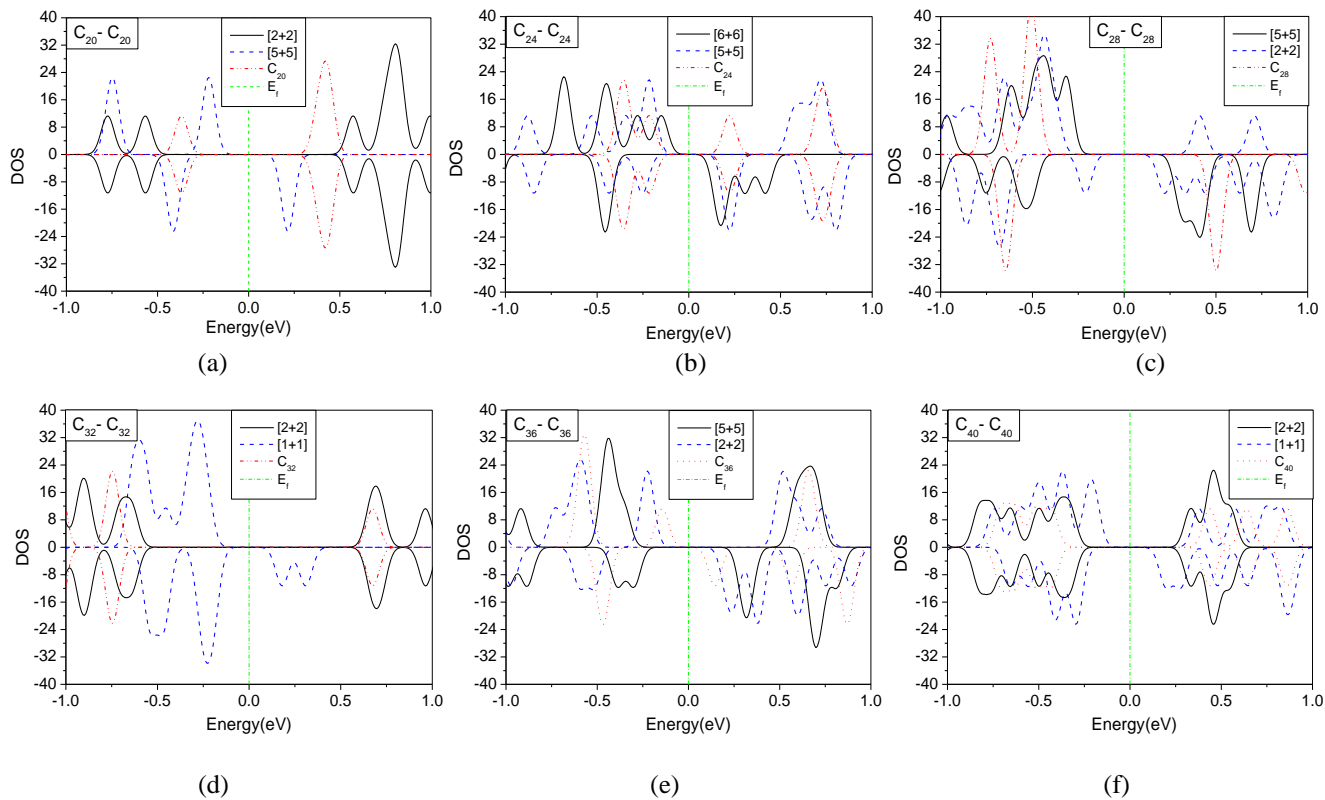

**Fig. 3**. DOS plots for symmetrical sized small fullerene dimers.

To understand the change in the electron density of dimer configuration w.r.t. individual fullerene, the Density of states (DOS) are calculated and shown in Fig. 3. Fig. 3(a) validates that $C_{20}$ fullerene cage and $C_{20}$-$C_{20}$ dimer in [2+2] mode are non-magnetic and [5+5] mode has spin polarized states. The origin of magnetic moment in [5+5] mode comes from the 2p orbitals because of unequal spin up and down states near the Fermi level. For $C_{24}$-$C_{24}$ dimer, [6+6] mode shows a significant change in DOS w.r.t. single $C_{24}$ cage which further points to its magnetic nature. [5+5] mode also show some decrease in the magnetization but there is a visible difference in up-down DOS (Fig. 3(b)). Both [6+6] and [5+5] modes show strong polarization around Fermi level in 2p orbitals while $C_{24}$ fullerene cage is non-magnetic and has equal up and down states. Similarly, the DOS for other dimer cages also show significant changes near Fermi level, which is due to the redistribution of electrons in 2p orbitals (see Fig. 3).

The symmetrical sized dimers have shown significant variation in their magnetic moment w.r.t. the type of inter-cage bonding. Their magnetic moments are mainly localized on first, second and third NNs from connecting bond atoms due to the redistribution of charges in spin up and spin down electron states. The contribution from connecting bond atoms towards total MM is very small which may be explained due to tetra-bonding of C atoms completing their valency by making four σ-bonds with neighboring C atoms. We have also plotted projected density of states in Fig 6 for few systems, which shows that 2p-orbitals in each case has maximum contribution towards total MM. The Muliken charge distribution analysis of the dimers suggest that the connecting bond C atoms losses charge in range 0.021 – 0.114 electrons, whereas the first NN gain charge of the order of 0.011 – 0.066 $e^-$s. However, the gain in charge for second NN is ~0.010 – 0.038 electrons. This redistribution of charges at different C sites is responsible for the variation in localized MMs. In symmetrical sized dimers, some of the C atoms show antiferromagnetic alignment w.r.t. their surrounding C atoms. However, the interaction between individual cages of the dimer is found to be ferromagnetic in nature.

### 3.2.2 Asymmetrical Sized Dimers;

The magnetic properties calculated for symmetrical sized dimers as described in the previous section were calculated for asymmetrical sized dimers and are tabulated in Table 5. Asymmetrical sized dimers with small difference in size were considered to understand their magnetic behavior. $C_{20}$ and $C_{24}$ in the isolated form are non-magnetic with zero magnetic moment, when they form $C_{20}$-$C_{24}$, the stable dimer structure with [2+2] mode remains non-magnetic. However, when $C_{20}$-$C_{24}$ connects with [1+1] and [5+5] bonding, the resultant structure becomes magnetic with magnetic moment of 1.9 $\mu_B$ and 2.0 $\mu_B$ respectively. The results suggest dependence of magnetic behavior on type of inter-connecting mode of dimer.

**Table-5.** Total Magnetic moments and HOMO-LUMO gaps for asymmetrical sized dimers.

| Dimer | Total Magnetic Moment ($\mu_B$) | | | | Dimer | HOMO-LUMO gaps (eV) | | | | | | | |
|---|---|---|---|---|---|---|---|---|---|---|---|---|---|
| $C_n$-$C_m$ | [1+1] | [2+2] | [5+5] | [6+6] | $C_n$-$C_m$ | [1+1] | | [2+2] | | [5+5] | | [6+6] | |
| | | | | | | ↑ | ↓ | ↑ | ↓ | ↑ | ↓ | ↑ | ↓ |
| $C_{20}$-$C_{24}$ | 1.90 | 0.00 | 2.00 | - | $C_{20}$-$C_{24}$ | 0.59 | 0.39 | 0.71 | 0.71 | 0.88 | 0.43 | - | - |
| $C_{24}$-$C_{28}$ | 2.00 | 2.00 | 4.00 | 3.95 | $C_{24}$-$C_{28}$ | 0.82 | 0.32 | 0.40 | 0.35 | 0.89 | 0.42 | 0.34 | 0.42 |
| $C_{28}$-$C_{32}$ | 4.00 | 2.00 | 4.00 | 2.00 | $C_{28}$-$C_{32}$ | 1.66 | 0.39 | 0.66 | 0.43 | 0.90 | 0.60 | 0.34 | 0.70 |
| $C_{32}$-$C_{36}$ | 1.95 | 2.00 | 2.00 | 4.00 | $C_{32}$-$C_{36}$ | 0.93 | 0.29 | 0.36 | 0.56 | 0.89 | 0.48 | 0.53 | 0.70 |
| $C_{36}$-$C_{40}$ | 2.00 | 3.96 | 2.00 | 4.00 | $C_{36}$-$C_{40}$ | 0.66 | 0.73 | 0.64 | 0.47 | 0.52 | 0.24 | 0.89 | 0.67 |
| $C_{40}$-$C_{44}$ | 2.00 | 4.00 | 2.00 | 0.00 | $C_{40}$-$C_{44}$ | 0.73 | 0.39 | 0.81 | 0.42 | 0.51 | 0.24 | 0.71 | 0.71 |
| $C_{44}$-$C_{50}$ | 1.98 | 2.00 | 1.92 | 0.00 | $C_{44}$-$C_{50}$ | 0.88 | 0.55 | 0.85 | 0.61 | 0.68 | 0.64 | 0.90 | 0.90 |

To understand the origin of magnetism in $C_{20}$-$C_{24}$, local magnetic moments on each C atom was calculated and are shown in Fig 4. The connecting bond atoms contribute very small magnetic moment towards total MM and the major contribution comes from second and third NNs from connecting bond which contribute 42% and 33% of total MM respectively. The HOMO-LUMO gap for spin up and spin down electron states show similar behavior as observed in symmetrical sized dimers.

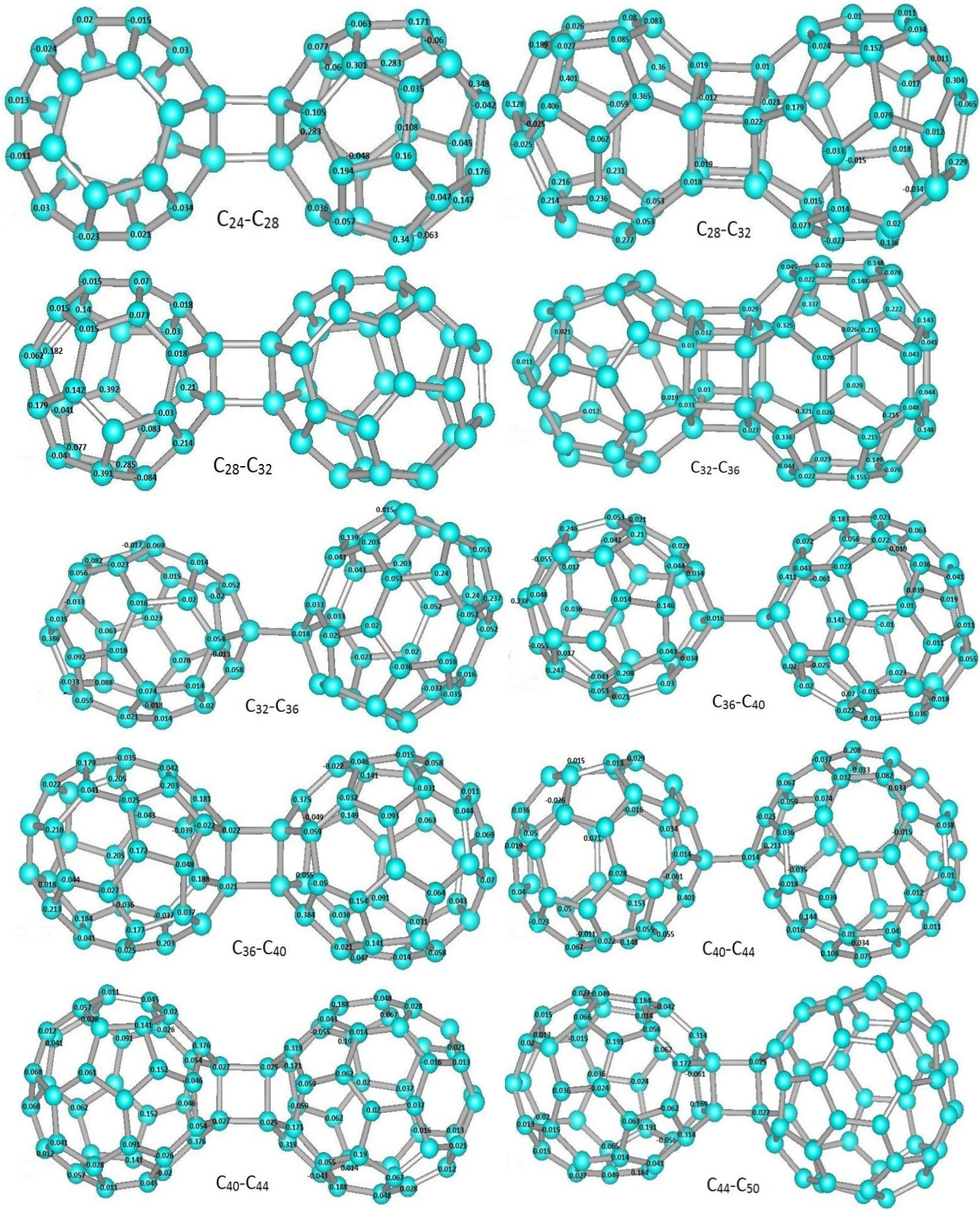

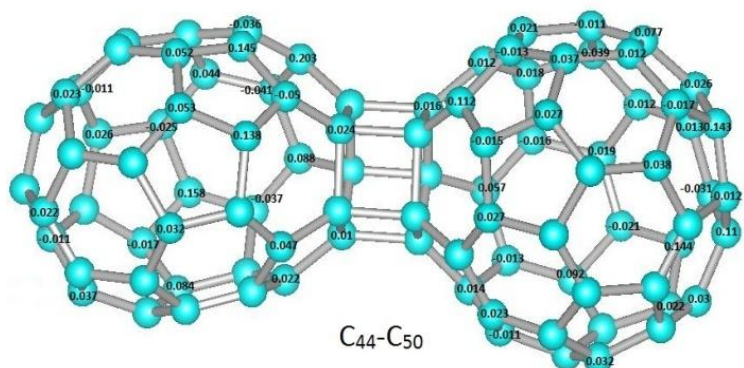

**Fig. 4**. Local magnetic moments of asymmetrical sized small fullerene dimers in $\mu_B$.

$C_{28}$ fullerene is magnetic with magnetic moment of 4 $\mu_B$ in isolated form. When it forms a dimer with $C_{24}$, the resultant most stable dimer structure with [5+5] mode becomes magnetic with total MM of 4.0 $\mu_B$. The total MM decreases in [1+1], [2+2] and [6+6] configurations having total MM 2.0 $\mu_B$, 2.0 $\mu_B$ and 3.95 $\mu_B$ respectively. The local MMs suggest that the maximum contribution of 70% comes from $C_{28}$ cage, while $C_{24}$ cage contributes only 30% of total MM, which indicates that $C_{28}$ is inducing magnetism in $C_{24}$. There is very small contribution of ~1.5% from connecting bond atoms towards total MM, whereas the maximum contribution comes from second and third NNs i.e. ~40% and 30% of total MM respectively. In [2+2] mode of $C_{24}$-$C_{28}$ dimer, only $C_{28}$ cage contributes towards total MM where local MM is distributed away from connecting bond atoms. The HOMO-LUMO gap for spin up and spin down states for all the modes show finite energy difference for magnetic dimer.

In the combination of $C_{28}$ and $C_{32}$ cages, the most stable dimer structure [2+2] becomes magnetic after dimerization with total MM of 2.0 $\mu_B$. The other modes [1+1], [5+5] and [6+6] of $C_{28}$-$C_{32}$ show magnetic behavior having total MM of 4.0 $\mu_B$, 4.0 $\mu_B$ and 2.0 $\mu_B$ respectively. The local MMs on each C atom were calculated to comment on the magnetic behavior of dimer structure. For most stable [2+2] mode, only $C_{28}$ cage contribute towards total MM as the cage is magnetic in isolated form and $C_{32}$ is non-magnetic. The magnetic moments are evenly distributed over first, third and fourth NNs from connecting bonds which contribute 23%, 32% and 36% respectively. The HOMO-LUMO gaps for spin up and down states have finite values.

In isolated form, $C_{32}$ fullerene is non-magnetic whereas $C_{36}$ has magnetic moment of 2.0 $\mu_B$. After dimerization, the resultant dimer structure in most stable [1+1] mode has total MM of 1.95 $\mu_B$. However, the total MM for [2+2], [5+5] and [6+6] increases to 2.0 $\mu_B$, 2.0 $\mu_B$ and 4.0 $\mu_B$ respectively. In ground state [1+1] mode, both the cages contribute equally towards total MM of the dimer and the magnetic moments are evenly distributed on third, fifth and seventh NNs from connecting bond atoms. The local MMs for [6+6] mode of $C_{32}$-$C_{36}$ dimer, contribution comes from $C_{36}$ cage only and contribution of ~36% comes from first and third NNs from connecting bonds. The HOMO-LUMO gap values for spin up and spin down states also show the magnetic behavior of dimer in all the configurations.

The combination of magnetic $C_{36}$ fullerene with non-magnetic $C_{40}$ fullerene results in a magnetic dimer structure having total MM of 2.0 $\mu_B$ in most stable [1+1] mode. In other configurations [2+2], [5+5] and [6+6], the total MM is 3.96 $\mu_B$, 2.0 $\mu_B$ and 4.0 $\mu_B$ respectively. The local MMs show that both the cages contribute equally towards total MM. The major contribution in total MM comes from first, third and fifth NNs from connecting bond atoms, whereas connecting bond atoms contribute ~1% towards total MM. The HOMO-LUMO gaps for spin up and spin down electron states show magnetic behavior of dimer and have finite values in all modes.

$C_{40}$ and $C_{44}$ cages are non magnetic in isolated form, but after dimerization the resultant dimer structure becomes magnetic in most stable [1+1] mode with total MM of 2.0 $\mu_B$. [2+2] and [6+6] modes of $C_{40}$-$C_{44}$ dimer are isoenergetic but they have different magnetic behavior. [6+6] mode is non-magnetic with zero magnetic moment, while [2+2] has high magnitude of total MM of order of 4.0 $\mu_B$. This explains the dependence of inter-cage bonding on the magnetic behavior of dimer formed. The local MMs suggest that each cage contributes ~50% towards total MM and the magnetic moments are not localized on connecting bond atoms. The main contribution comes from first and third NNs from connecting bonds with ~24% and ~20% respectively. The HOMO-LUMO gaps for spin up and down states are equal for [6+6] mode indicating its non-magnetic behavior.

**Fig. 5**. DOS plot for asymmetrical sized small fullerene dimers.

The non-magnetic $C_{44}$ and $C_{50}$ cages form a magnetic dimer structure with total MM of 2.0 $\mu_B$ in [1+1], [2+2] and [5+5] modes. However, the dimer in [6+6] mode remains non-magnetic. The local MMs for most stable [2+2] mode shows that only $C_{44}$ cage contribute towards total MM and it comes from first and third NNs with 48.5% and 41% respectively. The HOMO-LUMO gaps for spin up and spin down states in [1+1], [2+2] and [5+5] modes have finite energy difference which shows their magnetic behavior. The [6+6] configuration is non-magnetic having same spin up and down HOMO-LUMO gaps.

The density of state (DOS) plots have also been plotted in Fig. 5 to understand the magnetic behavior of asymmetrical sized dimers. The plots show that there is significant variation in spin up and down states near the Fermi level after dimerization of cages. In $C_{20}$-$C_{24}$ dimer, DOS plot give the magnetic behavior of [5+5] mode, as there are spin polarized states, while for [2+2] mode spin up and down states are identical (Fig. 5(a)). Similarly the DOS plots of other dimers show redistribution of electrons and presence of some unfilled states near the Fermi level and the contribution comes from 2p orbitals. The variation in spin up and down states of all these dimers points towards their magnetic nature.

**Table-6:** Contribution of 2s and 2p orbitals towards localized magnetic moment.

| Dimer | Local MM | 2s-orbital | 2p-orbital |
|---|---|---|---|
| $C_{24}$-$C_{28}$ | 0.413 | 0.028 | 0.387 |
| | 0.407 | 0.028 | 0.381 |
| | 0.363 | 0.023 | 0.341 |
| | 0.355 | 0.022 | 0.332 |
| | 0.278 | 0.017 | 0.264 |

Therefore, the results suggest that there is a large variation in magnetic behavior of individual cages when they form asymmetrical sized dimers. In some cases, both the cages contribute equally towards total MM, whereas in few of them only one cage contributes towards total MM. Further, the localized moments show that the connecting bond atoms are not the major contributors in total MM which may be due to the change in hybridization from $sp^2$ to $sp^3$. There is charge imbalance in spin up and spin down states of first, second and third NNs from connecting bonds which makes them contribute towards total MM. The local MMs show that some of the C atoms behave antiferromagnetically w.r.t. their neighboring C atoms. However, the individual cages in dimer structure interact ferromagnetically w.r.t. each other. Table 6 tabulates the local MM and contribution of 2s- and 2p-orbitals for five atoms of $C_{24}$-$C_{28}$ dimer starting from maximum value of local MM. The table shows that 2p-orbitals of C atom has major contribution to local MM, which further contributes to total MM of dimer structure. The plots of projected density of states (PDOS) also show the contribution of 2p-orbitals, which are plotted in Fig.6. The redistribution of charges on C atoms has been studied using Muliken charge analysis which shows that both the fullerene cages lose charge in the range $0.014 - 0.128$ $e^-$ for connecting bond atoms. The first NNs gain charge of the order of $0.012 - 0.076$ $e^-$s, whereas the gain in charge for second NNs is $0.002 - 0.029$ $e^-$s. This charge redistribution at different sites leads to variation in magnetic moments of individual C atoms.

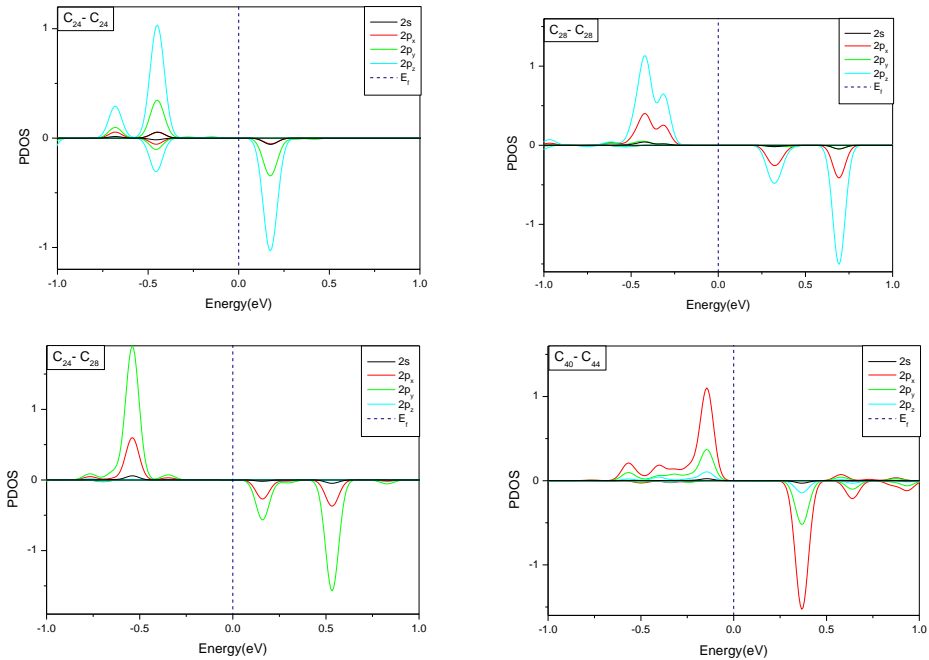

**Fig.6;** PDOS (projected density of states) plots for one atom of few symmetrical and asymmetrical sized small fullerene dimers.

## 4. CONCLUSIONS;

We have investigated a feasibility of formation of symmetrical and asymmetrical sized small fullerene dimers and their magnetic properties using spin polarized density functional theory. All possible modes ([1+1], [2+2], [5+5] and [6+6]) through which the cages can connect were considered. The negative values of interaction energies for all dimer combinations indicate strong possibility of their production except $C_{50}$-$C_{50}$ dimer which is unfavorable in their [1+1], [5+5] and [6+6] modes. All the dimer configurations are bonded weakly with connecting bonds ranging between 1.55 Å – 1.60 Å. The interaction energy suggests higher stability of symmetrical sized dimers than asymmetrical sized dimers.

Dimerization of the fullerenes result in significant change in electronic and magnetic properties w.r.t. isolated fullerene. All fullerenes in isolated form except $C_{28}$ and $C_{36}$ are non-magnetic and show interesting change in the magnetic behavior on dimerization. When two non-magnetic fullerenes are combined, the resultant dimer formed is magnetic. However, the combination of a magnetic and a non-magnetic fullerene leads to induced magnetism on non-magnetic fullerene in dimer configuration. When two magnetic fullerenes are combined, there is an enhancement or decrease in total magnetic moment of the resultant dimer. All magnetic dimers show different HOMO-LUMO gap for spin up and down electrons, whereas non-magnetic dimers have same HOMO-LUMO gap for spin up and spin down electrons. Magnitude of the magnetic moment is proportional to the difference in the HOMO-LUMO gap of spin up and spin down electron. The origin of magnetization may be understood in terms of structural distortion of fullerenes in dimer configuration and redistribution of charge from connecting C atoms due to formation of connecting bonds. More is the structural distortion more is the magnetic moment of the dimer structure. Further, given suitable experimental conditions, these small fullerene dimer structures can be produced, which can facilitate them as good magnetic materials having potential applications in spintronics.

## 5. ACKNOWLEDGEMENTS;

Authors are grateful to Siesta group for providing their computational code. Authors are thankful to DST-SERB, Govt. of India for providing financial support.